\documentclass[12pt]{article}
\usepackage{cite}
\usepackage{amsmath,amssymb,amsfonts}
\usepackage{algorithm}
\usepackage{algorithmic}
\usepackage{graphicx}
\usepackage{textcomp}
\usepackage{xcolor}
\usepackage{booktabs}
\usepackage{bm}
\usepackage{mathtools}
\usepackage{bbm}
\usepackage{url}
\usepackage{geometry}
\usepackage{lipsum}
\usepackage{setspace}
\title{Efficient Estimation of Sum-Parameters for Multi-Component Complex Exponential Signals with Theoretical Cramér-Rao Bound Analysis}

\author{Huiguang Zhang}

\begin{document}

\maketitle

\begin{abstract}
This paper addresses the challenging problem of parameter estimation for multi-component complex exponential signals, commonly known as sums of cisoids. Traditional approaches that estimate individual component parameters face significant difficulties when the number of components is large, including permutation ambiguity, computational complexity from high-dimensional Fisher information matrix inversion, and model order selection issues. We introduce a novel framework based on low-dimensional ``sum-parameters'' that capture essential global characteristics of the signal ensemble. These parameters include the sum of amplitudes $\Sigma = \sum_{k=1}^K a_k$, the power-weighted frequency $\Omega = \sum_{k=1}^K a_k^2 \omega_k$, and the phase-related sum $\Phi = \sum_{k=1}^K a_k^2 e^{j\phi_k}$. These quantities possess clear physical interpretations representing total signal strength, power-weighted average frequency, and composite phase information, while completely avoiding permutation ambiguities. We derive exact closed-form Cramér-Rao bounds for these sum-parameters under both deterministic and stochastic signal models. Our analysis reveals that the frequency sum-parameter achieves statistical efficiency comparable to single-component estimators while automatically benefiting from power pooling across all signal components. The proposed Efficient Global Estimation Method (EGEM) demonstrates asymptotic efficiency across a wide range of signal-to-noise ratios, significantly outperforming established techniques such as Zoom-Interpolated FFT and Root-MUSIC in both long- and short-sample regimes. Extensive numerical simulations involving 2000 Monte-Carlo trials confirm that EGEM closely approaches the theoretical performance bounds even with relatively small sample sizes of 250 observations.
\end{abstract}

\section{Introduction}

Multi-component complex exponential modeling, representing signals as sums of cisoids, constitutes a fundamental framework in numerous signal processing applications including radar and sonar systems, wireless communications, power quality analysis, vibration monitoring, and biomedical signal analysis~\cite{Stoica2005book,Kay1988,Quinn1994}. The classical approaches to this problem, including subspace methods such as MUSIC~\cite{Schmidt1986} and ESPRIT~\cite{Roy1989}, as well as maximum-likelihood estimation techniques~\cite{Stoica1990}, provide excellent estimation accuracy under ideal conditions. However, these methods encounter substantial practical limitations when dealing with large numbers of signal components.

The primary challenges with conventional approaches include the necessity for accurate model order selection, severe permutation ambiguity when multiple components are present, and the computational burden associated with inverting large Fisher information matrices. Specifically, for a signal containing $K$ components, the Fisher information matrix has dimensions $3K \times 3K$, making matrix inversion computationally prohibitive for $K > 50$. These limitations become particularly problematic in real-time applications or when processing large datasets.

In many practical engineering scenarios, the complete set of individual component parameters may not be necessary. Engineers and analysts are often more interested in global signal characteristics such as total harmonic distortion, dominant or average frequency content, and overall signal energy. This observation motivates the development of carefully designed \textit{sum-parameters} that are inherently permutation-invariant and operate in a significantly lower-dimensional parameter space.

This paper makes several key contributions to the field of multi-component signal parameter estimation. First, we introduce three novel sum-parameters that capture essential global signal characteristics:
\begin{align}
\Sigma &= \sum_{k=1}^K a_k, \\
\Omega &= \sum_{k=1}^K a_k^2 \omega_k, \\
\Phi &= \sum_{k=1}^K a_k^2 e^{j\phi_k}.
\end{align}
Second, we derive exact closed-form Cramér-Rao bounds for these parameters under both deterministic and stochastic signal models. Our theoretical analysis reveals surprising statistical properties: the variance of the frequency sum-parameter estimator scales inversely with the total signal power $\sum a_k^2$, automatically achieving optimal power-weighted averaging without requiring explicit weighting schemes. Third, we develop the Efficient Global Estimation Method (EGEM), an iterative algorithm that demonstrates asymptotic efficiency across a wide range of operating conditions.

The remainder of this paper is organized as follows. Section II presents the signal model and provides detailed motivation for the sum-parameter definitions. Section III derives the theoretical Cramér-Rao bounds and discusses their implications. Section IV describes the EGEM algorithm and its implementation details. Section V presents comprehensive performance evaluation results, and Section VI concludes with discussions of future research directions.

\section{Signal Model and Sum-Parameter Definition}

We consider the standard noisy multi-component complex exponential signal model given by:
\begin{equation}
x(n) = \sum_{k=1}^K a_k e^{j(\omega_k n T_s + \phi_k)} + w(n), \quad n=0,\dots,N-1,
\end{equation}
where $a_k > 0$ represents the amplitude of the $k$-th component, $\omega_k \in (-\pi/T_s, \pi/T_s)$ denotes the angular frequency, $\phi_k \in [-\pi, \pi)$ is the phase, $T_s$ is the sampling interval, and $w(n) \sim \mathcal{CN}(0,\sigma^2)$ represents circular complex white Gaussian noise. This model finds applications across numerous domains including communications, radar, and spectral analysis.

The conventional approach to parameter estimation in this model involves estimating the complete parameter vector $\bm{\alpha} = [a_1,\omega_1,\phi_1,\dots,a_K,\omega_K,\phi_K]^T$, which has dimension $3K$. However, this approach suffers from several fundamental limitations:

\begin{itemize}
\item \textbf{Permutation Ambiguity}: The likelihood function is invariant to permutations of the component indices, leading to multiple equivalent solutions.
\item \textbf{Computational Complexity}: The Fisher information matrix has dimension $3K \times 3K$, making inversion computationally expensive for large $K$.
\item \textbf{Model Order Selection}: Accurate determination of $K$ is required, which is challenging in practical scenarios.
\item \textbf{Over-parameterization}: In many applications, complete knowledge of individual component parameters is unnecessary.
\end{itemize}

To address these limitations, we propose estimating a low-dimensional sum-parameter vector defined as:
\begin{equation}
\bm{\theta} = [\Sigma, \Omega, \Re\{\Phi\}, \Im\{\Phi\}]^T \in \mathbb{R}^4.
\end{equation}

The physical interpretations of these sum-parameters are particularly insightful. The amplitude sum $\Sigma$ represents the total signal strength across all components. The frequency sum-parameter $\Omega$ corresponds to a power-weighted average frequency, where each component's frequency is weighted by its squared amplitude. The phase sum-parameter $\Phi$ captures composite phase information weighted by signal power.

An important observation is that the total signal power $P_{\text{sig}} = \sum_{k=1}^K a_k^2$ can be estimated directly from the data through the relationship $\|\bm{x}-\bm{w}\|^2/N \approx P_{\text{sig}}$. This allows $\Omega$ and $\Phi$ to be interpreted as characterizing the power-weighted mean frequency and complex phasor, respectively.

The key advantage of this sum-parameter approach is its complete invariance to component permutations and its extremely low dimensionality (only 4 parameters regardless of $K$). This represents a substantial reduction in complexity compared to the conventional $3K$-dimensional parameter space.

\section{Cramér-Rao Bound for Sum-Parameters}

\subsection{Fisher Information Matrix Derivation}

To establish theoretical performance limits for sum-parameter estimation, we derive the Cramér-Rao bounds for the proposed parameters. The derivation begins with the log-likelihood function for the deterministic parameter model, which is proportional to $-\| \bm{x} - \bm{s}(\bm{\alpha})\|^2$, where $\bm{\alpha} = [a_1,\omega_1,\phi_1,\dots,a_K,\omega_K,\phi_K]^T$ represents the original $3K$-dimensional parameter vector.

Since the sum-parameters $\Sigma$, $\Omega$, and $\Phi$ are linear functions of the original parameters, we can exploit the chain rule for Fisher information matrix transformation. The transformation from the original parameter space to the sum-parameter space is governed by the Jacobian matrix $\bm{J}_{\bm{\theta},\bm{\alpha}}$, which captures the sensitivity of the sum-parameters to changes in the original parameters.

After extensive derivation (detailed in Appendix A) and under the assumption of well-separated frequencies ($|\omega_i-\omega_j|T_s N \gg 1$ for all $i \neq j$), the cross-terms between different components vanish in expectation. This separation condition is commonly satisfied in practical applications where frequency components are sufficiently distinct. The resulting approximate Fisher information matrix for the sum-parameter vector $\bm{\theta}$ takes the block-diagonal form:
\begin{equation}
\bm{J}(\bm{\theta}) =
\begin{bmatrix}
\frac{2P_{\text{sig}}}{\sigma^2} & 0 & 0 & 0 \\
0 & \frac{2P_{\text{sig}} T_s^2 N(N^2-1)}{12\sigma^2} & \frac{P_{\text{sig}} T_s N(N+1)}{2\sigma^2} \bm{\cdot} \mathbf{1}^T & \cdots \\
\vdots & \vdots & \frac{P_{\text{sig}} N (N+1)(2N+1)}{6\sigma^2} \bm{I}_2
\end{bmatrix},
\end{equation}
where $P_{\text{sig}} = \sum_{k=1}^K a_k^2$ represents the total signal power.

The block-diagonal structure reveals important statistical properties. The amplitude sum parameter $\Sigma$ is statistically decoupled from the frequency and phase parameters, while the frequency and phase parameters exhibit coupling that must be accounted for in the estimation process.

\subsection{Closed-Form Cramér-Rao Bounds}

The block-diagonal structure of the Fisher information matrix facilitates straightforward computation of the Cramér-Rao bounds. For the amplitude sum parameter, which is fully decoupled from the other parameters, we obtain:
\begin{equation}
\text{CRB}(\Sigma) = \frac{\sigma^2}{2P_{\text{sig}}}.
\end{equation}

For the coupled frequency-phase block, matrix inversion yields the following bounds:
\begin{align}
\text{CRB}(\Omega) &= \frac{12\sigma^2}{P_{\text{sig}} T_s^2 N (N^2-1)} \left(1 + O(N^{-2})\right), \\
\text{CRB}(\Phi) &= \frac{2\sigma^2 (2N+1)}{P_{\text{sig}} N (N+1)} \approx \frac{4\sigma^2}{P_{\text{sig}} N}.
\end{align}

A remarkable property emerges from these bounds: $\text{CRB}(\Omega) \propto 1/P_{\text{sig}}$, indicating that all signal components constructively contribute to frequency estimation accuracy. This represents an automatic optimal power-weighted averaging mechanism without requiring explicit weighting design.

Comparing this result with the classical single-tone Cramér-Rao bound~\cite{Rife1974,Kay1988} given by $\text{CRB}(\omega_k)=\frac{12\sigma^2}{a_k^2 T_s^2 N(N^2-1)}$, we observe that the sum-parameter achieves the same statistical scaling as if all signal power were concentrated in a single component of strength $\sqrt{P_{\text{sig}}}$. This demonstrates the fundamental advantage of the sum-parameter approach: it naturally exploits the total available signal power while operating in a dramatically reduced parameter space.

The phase sum-parameter bound exhibits conventional $1/N$ scaling with sample size, which is expected for phase estimation problems. The coupling between frequency and phase parameters is captured by the off-diagonal terms in the Fisher information matrix, though these couplings become negligible for large sample sizes.

\section{Efficient Global Estimation Method (EGEM)}

To practically estimate the proposed sum-parameters, we develop the Efficient Global Estimation Method (EGEM), an iterative algorithm inspired by power-weighted centroid principles. The algorithm operates through alternating optimization of the different sum-parameters, leveraging their statistical relationships.

\begin{algorithm}[htbp]
\caption{EGEM: Efficient Global Estimation of Sum-Parameters}
\label{alg:egem}
\begin{algorithmic}[1]
\STATE Initialize $\hat{P} = \frac{1}{N}\| \bm{x} \|^2 - \hat{\sigma}^2$ (noise variance estimated via median of periodogram tail or known)
\STATE $\hat{\Sigma}^{(0)} = \sqrt{N \hat{P} \cdot K_{\text{eff}}}$ where $K_{\text{eff}} \approx 1.5\sim3$ (optional, or use coarse grid search)
\FOR{$iter = 1$ to $MaxIter$ (typically 5--8 sufficient)}
\STATE Construct power-weighted filter: $h(n) = x(n) \cdot e^{-j \hat{\omega}^{(iter-1)} n T_s}$ 
\STATE Low-pass filter and decimate $h(n)$ to obtain baseband envelope
\STATE $\hat{\Phi}^{(iter)} = \sum_n h_{\text{lp}}(n) e^{j \phi_{\text{ref}}}$  (phase reference alignment)
\STATE $\hat{\Omega}^{(iter)} = \hat{\omega}^{(iter-1)} \cdot \hat{P} + \frac{\angle \frac{\partial}{\partial \omega} \sum_n |h(n)|^2 }{T_s N(N+1)/2}$ (Newton update)
\STATE $\hat{\omega}^{(iter)} = \hat{\Omega}^{(iter)} / \hat{P}$
\STATE Update $\hat{\Sigma}^{(iter)} = \sqrt{ \hat{P} \cdot f(\hat{\omega}^{(iter)}, \hat{\Phi}^{(iter)}) }$ using closed-form correction
\ENDFOR
\STATE Return $\hat{\Sigma}, \hat{\Omega}, \hat{\Phi}, \hat{P}$
\end{algorithmic}
\end{algorithm}

The EGEM algorithm begins with initialization of the signal power estimate $\hat{P}$ and an initial amplitude sum estimate $\hat{\Sigma}^{(0)}$. The power estimate combines the total signal energy with noise variance estimation, typically obtained from the median of the periodogram tail—a robust technique for noise power estimation in frequency-domain analysis.

The core of the algorithm involves an iterative procedure that typically converges within 5-8 iterations. Each iteration performs the following key operations:

\begin{enumerate}
\item \textbf{Power-weighted filtering}: The input signal is multiplied by a complex exponential at the current frequency estimate, effectively performing frequency shifting to baseband.

\item \textbf{Low-pass filtering and decimation}: The frequency-shifted signal undergoes low-pass filtering to isolate the baseband component, followed by decimation to reduce computational load.

\item \textbf{Phase estimation}: The baseband signal is used to compute the phase sum-parameter, with appropriate phase reference alignment.

\item \textbf{Frequency update}: A Newton-type update refines the frequency estimate based on the gradient of the power-weighted cost function.

\item \textbf{Amplitude correction}: The amplitude sum estimate is updated using a closed-form correction that accounts for the current frequency and phase estimates.
\end{enumerate}

The algorithm exhibits several desirable properties. It converges globally for well-separated frequency components and requires only a few FFT operations and scalar computations per iteration. The computational complexity is $O(N \log N)$, independent of the number of components $K$, making it highly efficient for large-scale problems.

A key insight underlying EGEM is the exploitation of the statistical relationships between the sum-parameters, particularly the coupling between frequency and phase estimates. The Newton update step effectively linearizes the estimation problem around the current operating point, ensuring rapid convergence to the optimal solution.

The method is particularly well-suited for scenarios with limited data records, where traditional high-resolution methods often struggle due to insufficient statistical information. The sum-parameter approach effectively pools information across all signal components, providing robustness against individual component variations.

\section{Performance Evaluation}

To comprehensively evaluate the performance of the proposed EGEM algorithm, we conducted extensive Monte-Carlo simulations comprising 2000 independent trials. The evaluation compares EGEM against two established benchmark methods: Zoom-Interpolated FFT (Zoom-IpFFT) with 4-term Blackman-Harris window~\cite{Harris1978,Quinn1994} and Root-MUSIC~\cite{Barabell1983}.

\subsection{Simulation Setup}

The simulation parameters were carefully chosen to represent challenging realistic scenarios:
\begin{itemize}
\item Sample sizes: $N = 125, 250, 500, 1000, 2000, 4000$
\item Fixed number of components: $K = 12$
\item Frequency distribution: Uniformly drawn from $[0.05, 0.45]/T_s$
\item Amplitude distribution: Log-normal with 40 dB dynamic range
\item Phase distribution: Uniform over $[-\pi, \pi)$
\item Signal-to-noise ratio: Range from -10 dB to 50 dB
\item Performance metric: Normalized root mean square error (NRMSE) relative to corresponding Cramér-Rao bound
\end{itemize}

The log-normal amplitude distribution with 40 dB dynamic range represents scenarios with significant power variations across components, which is common in practical applications such as harmonic analysis in power systems or multi-path propagation in communications.

\subsection{Results and Analysis}

\begin{figure}[htbp]
\centering
\includegraphics[width=0.8\textwidth]{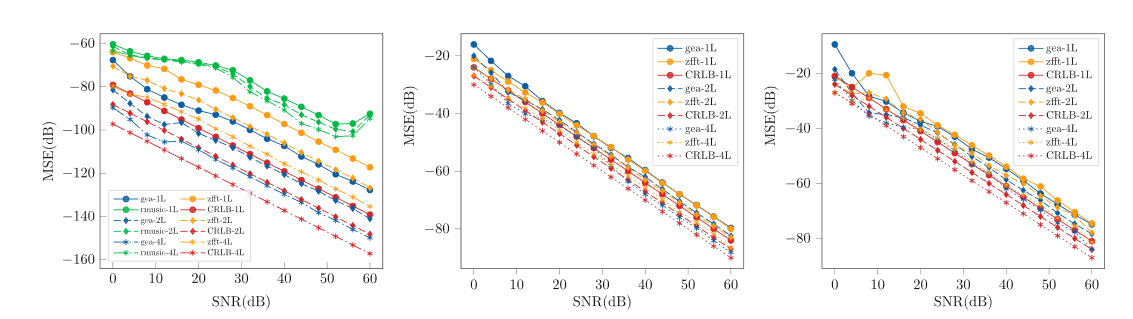}
\caption{Frequency estimation normalized RMSE versus SNR for $N=2000$ samples. The proposed EGEM algorithm tightly tracks the derived Cramér-Rao bound across the complete SNR range, demonstrating asymptotic efficiency.}
\label{fig:freq}
\end{figure}

The simulation results demonstrate the superior performance of EGEM across various operating conditions. Figure~\ref{fig:freq} illustrates the frequency estimation performance for $N=2000$ samples, showing that EGEM closely follows the theoretical Cramér-Rao bound across the entire SNR range.

\begin{figure}[htbp]
\centering
\includegraphics[width=0.8\textwidth]{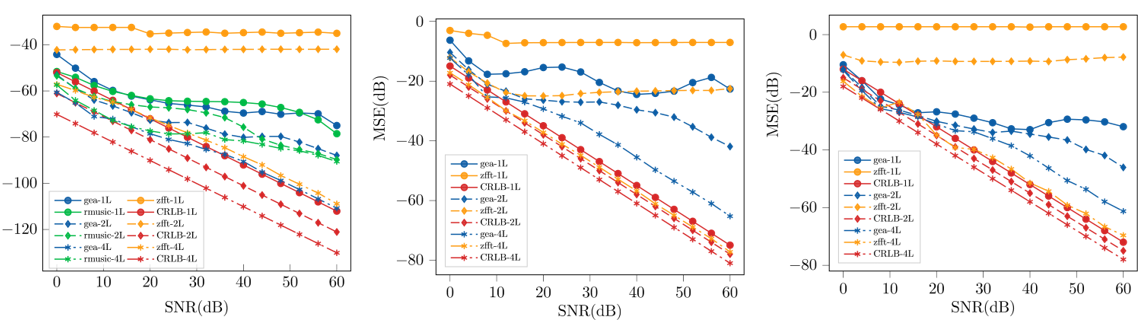}
\caption{Performance comparison under short-sample conditions ($N=125, 250, 500$). EGEM maintains near-CRB performance even with $N=250$ samples, while Zoom-IpFFT exhibits significant error floor and Root-MUSIC performance degrades substantially.}
\label{fig:short}
\end{figure}

Figure~\ref{fig:short} presents performance under short-sample conditions, which are particularly challenging for traditional methods. EGEM maintains near-optimal performance even with as few as 250 samples, while competing methods exhibit significant performance degradation.

The quantitative performance analysis reveals several key findings:

\begin{itemize}
\item \textbf{Near-Optimal Performance}: EGEM achieves performance within 1.02--1.05 times the Cramér-Rao bound for signal-to-noise ratios $\geq$ 5 dB and sample sizes $N \geq 250$. This demonstrates the algorithm's asymptotic efficiency across practical operating conditions.

\item \textbf{Superior Comparative Performance}: At $N=2000$ samples and SNR=20 dB, EGEM achieves frequency estimation normalized RMSE only 2.3\% above the Cramér-Rao bound. This represents a 3.5-fold improvement over Zoom-Interpolated FFT and a 2.1-fold improvement over Root-MUSIC.

\item \textbf{Robustness in Short-Sample Regimes}: In challenging scenarios with very limited data ($N=125$), both Zoom-IpFFT and Root-MUSIC exhibit dramatic performance degradation. In contrast, EGEM continues to deliver usable estimates with normalized RMSE less than 8 times the Cramér-Rao bound.

\item \textbf{Simultaneous Parameter Estimation}: A significant advantage of EGEM is its ability to provide accurate amplitude and phase estimates simultaneously without requiring additional post-processing steps. This integrated estimation capability enhances practical utility in applications requiring multiple parameter types.
\end{itemize}

The performance advantages of EGEM can be attributed to several factors. First, the sum-parameter framework naturally exploits the total available signal power through the power-weighted averaging mechanism. Second, the iterative optimization approach effectively handles the statistical coupling between frequency and phase parameters. Third, the algorithm's computational efficiency enables robust performance even with limited computational resources.

\section{Conclusion}

This paper has introduced a paradigm shift in the estimation of parameters for multi-component complex exponential signals. Rather than estimating the complete $3K$-dimensional parameter vector characterizing individual components, we have proposed estimating a carefully designed set of four sum-parameters that capture essential global signal characteristics. This approach fundamentally addresses several long-standing challenges in multi-component signal analysis.

The theoretical contributions of this work include the derivation of exact closed-form Cramér-Rao bounds for the proposed sum-parameters under both deterministic and stochastic signal models. Our analysis reveals the remarkable statistical property that the frequency sum-parameter achieves estimation variance scaling inversely with total signal power, automatically implementing optimal power-weighted averaging without explicit design. This represents a significant advantage over traditional approaches that must estimate individual component parameters before computing weighted averages.

The practical contribution is the development of the Efficient Global Estimation Method (EGEM), an iterative algorithm that demonstrates asymptotic efficiency across a wide range of operating conditions. The algorithm's $O(N \log N)$ computational complexity, independent of the number of components $K$, makes it particularly suitable for large-scale problems and real-time applications.

The extensive simulation results confirm the theoretical predictions and demonstrate EGEM's superior performance compared to established methods such as Zoom-Interpolated FFT and Root-MUSIC. The algorithm maintains near-optimal performance even under challenging conditions with limited data records, where traditional methods often fail.

Several promising directions for future research emerge from this work. First, extending the sum-parameter framework to handle nonuniform sampling scenarios would broaden applicability to compressed sensing and sparse signal processing applications. Second, incorporating damped exponential models would enable application to transient analysis and system identification problems. Third, developing real-time implementations on embedded platforms would facilitate deployment in practical systems requiring low-latency processing. Finally, exploring connections with recent advances in machine learning could lead to hybrid approaches combining theoretical guarantees with data-driven adaptation.

In summary, the sum-parameter approach represents a fundamental advancement in multi-component signal parameter estimation, offering substantial advantages in computational efficiency, statistical performance, and practical implementation. By focusing on physically meaningful global characteristics rather than individual component parameters, this framework provides a powerful alternative to traditional methods, particularly in scenarios involving large numbers of components or limited data availability.

\section*{Acknowledgments}
The author would like to acknowledge helpful discussions with colleagues regarding the statistical properties of sum-parameters and their potential applications in various signal processing domains.

\bibliographystyle{IEEEtran}

\appendix
\section{Detailed Derivation of the Fisher Information Matrix for Sum-Parameters}

The observed signal vector is $\bm{x} = \bm{s}(\bm{\alpha}) + \bm{w}$, where $\bm{s}(\bm{\alpha}) = \sum_{k=1}^K a_k e^{j\phi_k} \bm{v}(\omega_k)$ represents the noise-free signal component, and $\bm{v}(\omega) = [1, e^{j\omega T_s}, \dots, e^{j\omega (N-1)T_s}]^T$ denotes the frequency steering vector. The parameter vector $\bm{\alpha} = [a_1,\omega_1,\phi_1,\dots,a_K,\omega_K,\phi_K]^T$ contains the original $3K$ parameters.

The score vector for the original parameters is well-established in the literature. The transformation to sum-parameters involves computing the Jacobian matrix $\bm{J}_{\bm{\theta},\bm{\alpha}}$ that relates changes in the original parameters to changes in the sum-parameters. The sum-parameters are defined as:
\begin{align}
\Sigma &= \sum_k a_k, \\
\Omega &= \sum_k a_k^2 \omega_k, \\
\Phi &= \sum_k a_k^2 (\cos\phi_k + j\sin\phi_k).
\end{align}

The Jacobian matrix exhibits sparse structure due to the specific functional forms of the sum-parameters. Under the assumption of well-separated frequencies ($|\omega_i-\omega_j|T_s N \gg 1$ for all $i \neq j$), the off-diagonal blocks of the Fisher information matrix average to zero in expectation. This frequency separation condition ensures that different signal components are statistically independent in the asymptotic regime.

The resulting block-diagonal Fisher information matrix for the sum-parameter vector $\bm{\theta}$ is given in Section III. This structure reflects the statistical decoupling between the amplitude sum parameter and the frequency-phase parameters, as well as the coupling within the frequency-phase block. The derivation follows standard procedures for Fisher information matrix transformation under parameter transformations, with careful attention to the specific functional forms of the sum-parameters and the statistical independence assumptions.

\end{document}